\documentclass{mn2e}

\input epsf

\usepackage{graphicx}
\usepackage{txfonts}
\usepackage{epsfig}

\title[Distortion of p-mode peak profiles by the solar
cycle]{Distortion of the p-mode peak profiles by the solar-cycle
frequency shifts: do we need to worry?}

\author[W.~J.~Chaplin et al]{ W.~J.~Chaplin,$^1$ Y.~Elsworth,$^1$
R.~New,$^2$ and T.~Toutain,$^1$ \\$^1$ School of Physics and
Astronomy, The University of Birmingham, Edgbaston, Birmingham B15
2TT, UK \\$^2$Faculty of Arts, Computing, Engineering and Sciences,
Sheffield Hallam University, Sheffield S1 1WB, U.K.}

\begin{document}

\label{firstpage}

\maketitle

\begin{abstract}

We seek to address whether solar-cycle frequency shifts of the Sun's
low-$l$ p modes `distort' the underlying shapes of the mode peaks,
when those peaks are observed in power frequency spectra made from
data spanning large fractions, or more, of the cycle period. We
present analytical descriptions of the expected profiles, and validate
the predictions through use of artificial seismic timeseries data, in
which temporal variations of the oscillator frequencies are
introduced. Our main finding is that for the Sun-like frequency shifts
the distortion of the asymmetrical Lorentzian-like profiles is very
small, but also just detectible. Our analysis indicates that by
fitting modes to the usual Lorentzian-like models -- which do not
allow for the distortion -- rather than new models we derive, there is
a bias in the mode height and linewidth parameters that is comparable
in size to the observational uncertainties given by multi-year
datasets. Bias in the frequency parameter gives much less cause for
worry, being over an order of magnitude smaller than the corresponding
frequency uncertainties. The distortion discussed in this paper may
need to be considered when multi-year Sun-like asteroseismic datasets
are analyzed on stars showing strong activity cycles.

\end{abstract}


\maketitle

\section{Introduction}
\label{sec:intro}

High-quality observations of the solar p modes are now available,
which for some instruments cover almost three complete 11-yr cycles of
activity (Chaplin et al. 2007a).  Accurate and precise mode parameter
data are a vital prerequisite for making accurate inference on the
solar interior, be that on the hydrostatic or dynamic structure. There
are clear advantages to be gained by extracting estimates of the mode
parameters from power frequency spectra made from several, sometimes
many, years of contiguous observations. The excellent resolution in
frequency, and excellent height-to-background ratios observed in the
mode peaks, then allows mode parameters to be extracted to extremely
high precision. Subtle phenomena, such as asymmetry of mode peaks (a
diagnostic of the stochastic excitation, and granulation) and
asymmetry of mode frequency splittings (a diagnostic of the surface
activity) may then also be extracted reliably from the data. And the
weakly damped p modes at very low frequencies become visible and
amenable to measurement and study.

The observations may then span a sizeable fraction, or more, of an
11-yr solar activity cycle period. The question arises: what effect do
the well-known solar-cycle shifts in frequency through a long
timeseries have on the underlying shapes of the mode peaks, when those
peaks are observed in power frequency spectra made from the full
timeseries? Are the shapes so \emph{distorted} from the
Lorentzian-like form that the peak-bagging codes -- with their
Lorentzian-like fitting models -- inevitably return biased estimates
of the parameters? In this paper, we seek to answer this question for
the low-degree (low-$l$) core-penetrating solar p modes. We seek an
analytical description of the underlying peak-shapes expected from
timeseries in which the frequencies of modes are known to vary. We
then look at whether the form given is significantly different from
the assumed Lorentzian-like profiles. We also use simulations of
artificial timeseries data, in which temporal variations of the
oscillator frequencies have been introduced, to validate use of the
analytical expressions.

\begin{figure*}
 \centerline{\epsfxsize=8cm\epsfbox{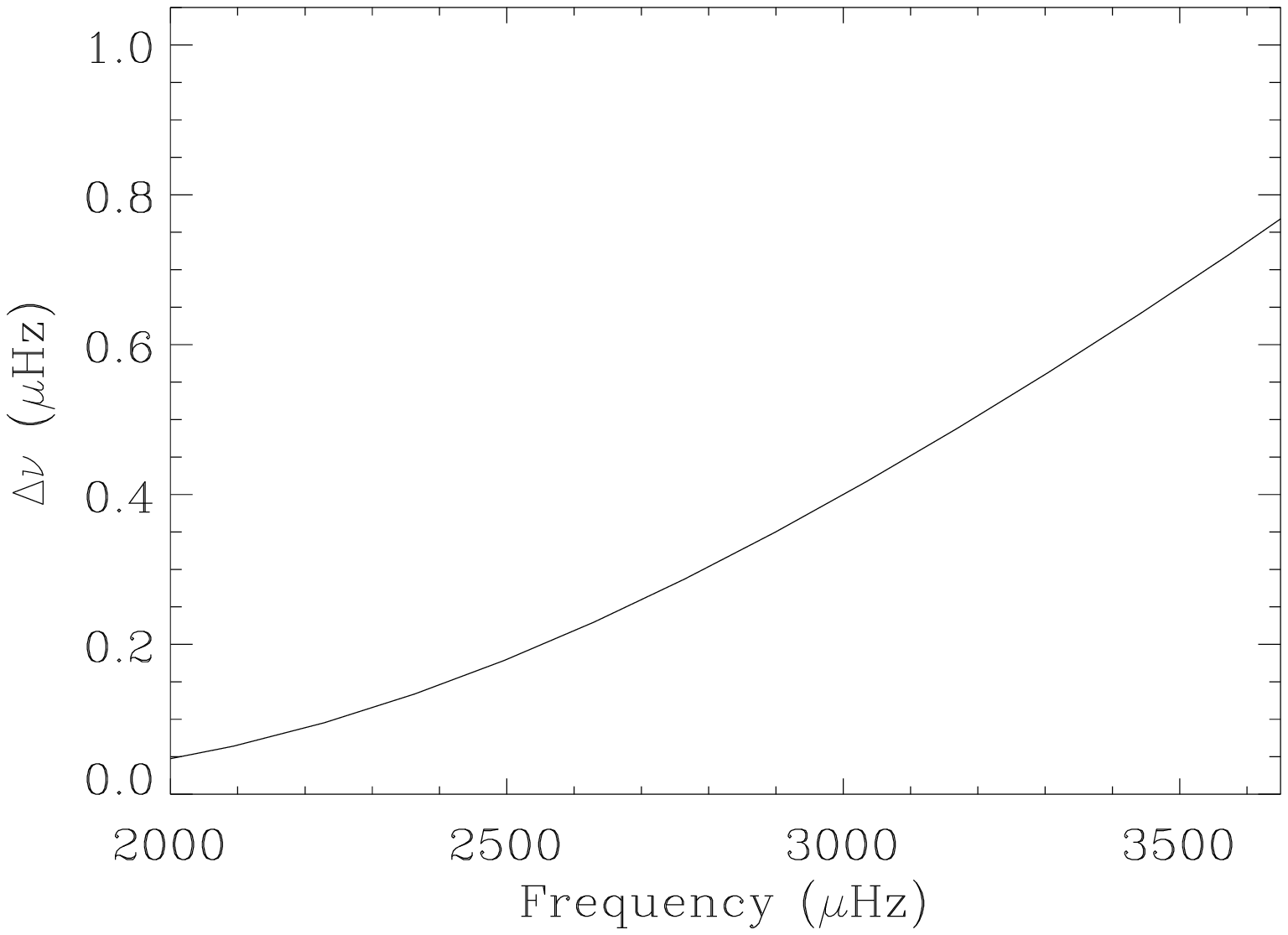}
 \quad
 \epsfxsize=8cm\epsfbox{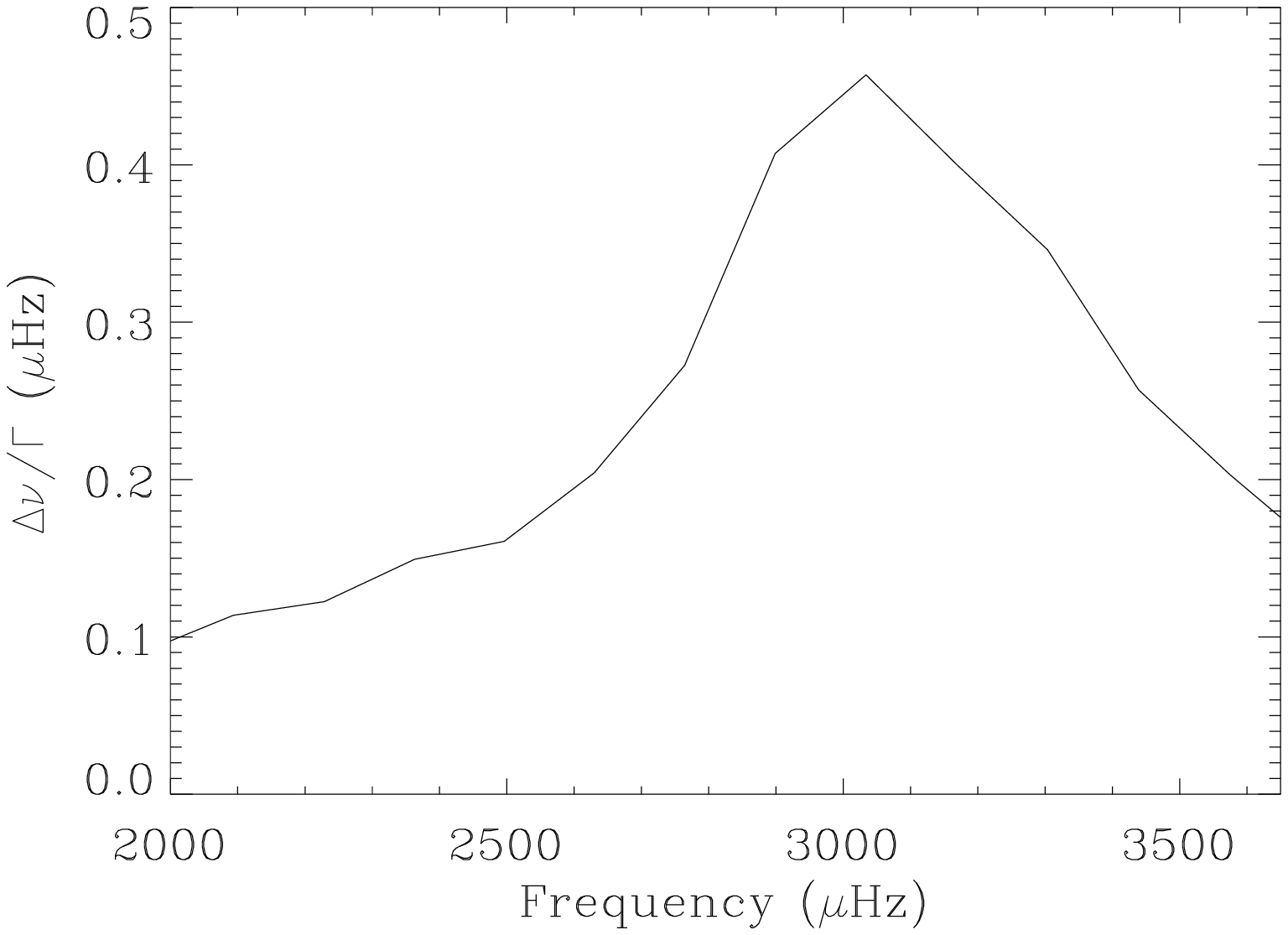}}

 \caption{Left-hand panel: smoothed representation of the solar activity cycle
 frequency shifts (from minimum to maximum activity), averaged across
 $l=0$ to 3. Right-hand panel: Ratio of the shifts plotted in the
 left-hand panel and the linewidths of the modes.}

 \label{fig:dnu}
 \end{figure*}

The frequency-shift regime of interest is illustrated in
figure~\ref{fig:dnu}. The left-hand panel plots the sizes of the
minimum-to-maximum solar-cycle frequency shifts of the low-$l$ modes
(here, averaged over $l=0$ to 3), as a function of mode frequency
(e.g., see Chaplin 2004). The average shift amounts to about $0.4\,\rm
\mu Hz$ for a mode at $\sim 3000\, \rm \mu Hz$. While modes at higher
frequency suffer a bigger shift (e.g., about $1\,\rm \mu Hz$ at
frequency $4000\, \rm \mu Hz$) their linewidths are then also about an
order of magnitude larger than at $3000\, \rm \mu Hz$, and it is the
shift-to-linewidth ratio that is of relevance for determining the
impact of any distortion. This ratio is plotted in the right-hand
panel of figure~\ref{fig:dnu}: clearly, modes at the centre of the
low-$l$ p-mode spectrum are potentially most susceptible to the
distortion effect.


 \section{Model of mode profile for frequency-shifted data}

To obtain a model of the underlying profile of a p mode whose
frequency is shifted in time, we note that the modes are excited and
damped on a timescale much shorter than that on which any significant
change of their frequency is observed. We therefore assume the
resultant profile corresponds to the average of all the instantaneous
profiles taken at any time $t$ within the full period of observation
$T$. We shall initially model the mode profiles as simple Lorentzians.
Each instantaneous profile is therefore described as a Lorentzian with
a central frequency $\nu(t)$, i.e., the `mode frequency' at time
$t$. The time-averaged profile is then
 \begin{equation}
 \label{eq:meanprofile}
 \left< P(\nu) \right> = \frac{1}{T}
 \int_{0}^{T} \frac{H}{1 + \left( \frac{\nu - \nu(t)}{\Gamma/2}
 \right)^2}~dt,
 \end{equation}
where the angled brackets indicate an average over time, and $H$ and
$\Gamma$ are the mode height (maximum power spectral density) and
linewidth, respectively.

In what follows we shall discuss the profiles given for two functions
describing the frequency shifts in time: first, the simplest possible
function, this being a linear variation over time; and second a
cosinusoidal variation to mimic the full solar activity cycle. We have
neglected the effects of the solar-cycle variations in height and
width, which will be included in future work.

\section{Analytical mode profiles}
\label{sec:analyt}
 
\subsection{Linear variation in time}
\label{sec:lin}

\begin{figure}
 \centerline{\epsfxsize=8cm\epsfbox{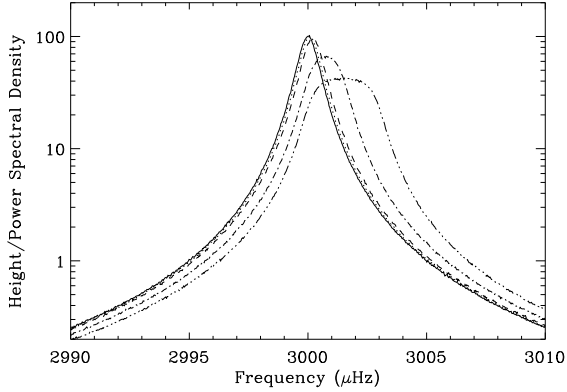}}

 \caption{Peak profiles expected for a single mode of width $1\,\rm
 \mu Hz$ in the power frequency spectrum of a time series within which
 the frequency was varied in a linear manner by total amount
 $\Delta\nu$.  Various linestyles are: no shift (solid line);
 $\Delta\nu=0.15\,\rm \mu Hz$ (dotted line); $0.40\,\rm \mu Hz$
 (dashed line); $1.50\,\rm \mu Hz$ (dot-dashed line); and $3.0\,\rm
 \mu Hz$ (dot-dot-dot-dashed line).}

 \label{fig:profs}
 \end{figure}

We begin by assuming a simple linear variation in time, i.e.,
 \begin{equation}
 \label{eq:modefreq}
 \nu(t) = {\nu_0} + \Delta \nu \frac{t}{T},
 \label{eq:lin}
 \end{equation}
where $\nu_0$ is the unperturbed frequency, and the shift is
calibrated so that the total shift from the start ($t=0$) to the end
($t=T$) of the timeseries is $\Delta\nu$. Substitution of
equation~\ref{eq:lin} into equation~\ref{eq:meanprofile}, followed by
solution of the integral, gives the predicted mode profile:
 \begin{equation}
 \label{eq:linmeanprofile}
 \left<P(\nu) \right> = \frac{H}{2\epsilon} {\rm atan} \left( \frac{2 \epsilon}{1 - \epsilon^2 +
 X^2}\right),
 \end{equation}
where
 \begin{equation}
 \label{eq:epsi}
 \epsilon = \frac{ \Delta \nu} {\Gamma}
 \end{equation}
is the frequency shift in units of the mode linewidth and
 \begin{equation}
 \label{eq:X}
 X= \frac{\nu - (\nu_0+ \Delta\nu/2)}{\Gamma/2}.
 \end{equation}
Figure~\ref{fig:profs} shows profiles given by
equation~\ref{eq:linmeanprofile}. The unperturbed profile (solid line)
is for a mode having an unperturbed frequency of $\nu_0=3000\, \rm \mu
Hz$, an unperturbed linewidth of $\Gamma = 1\, \rm \mu Hz$, and an
unperturbed height of $H=100$ units. The other curves show the
profiles that result when the frequency shift, $\Delta\nu$, is:
$0.15\,\rm \mu Hz$ (dotted line); $0.40\,\rm \mu Hz$ (dashed line);
$1.50\,\rm \mu Hz$ (dot-dashed line); and $3.0\,\rm \mu Hz$
(dot-dot-dot-dashed line). Since $\Gamma = 1\, \rm \mu Hz$, the
$\Delta \nu$ also correspond to the shift-to-linewidth ratios,
$\epsilon$. To put the values in context, low-$l$ modes at $\approx
3000\, \rm \mu Hz$, which also have width $\approx 1\,\rm \mu Hz$,
suffer a total fractional shift of approximately $0.40\,\rm \mu Hz$
from the minimum to the maximum of the solar activity cycle.

With reference to figure~\ref{fig:profs}, it is evident that only at
the two largest shifts (dot-dashed and dot-dot-dot-dashed lines) do
the profiles depart appreciably from the Lorentzian form. However,
these shifts are somewhat larger than those suffered by the real p
modes. Closer inspection of the profiles does reveal some modest
distortion at the two, smaller, Sun-like shifts. These have $\epsilon
= 0.15$ and 0.40 respectively. We discuss the implications of this
distortion for introducing bias in results of the usual peak fitting
in Section~\ref{sec:bias} below.

We also tested the predictions of equation~\ref{eq:linmeanprofile}
with Monte Carlo simulations of artificial data. The Laplace transform
solution of the equation of a forced, damped harmonic oscillator was
used to generate mode components at a 40-s cadence in the time domain,
in the manner described by Chaplin et al. (1997).  Components were
re-excited independently at each sample with small `kicks' drawn from
a Gaussian distribution. The simulations gave modes having Lorentzian
limit shapes in the power frequency spectrum. The underlying frequency
of the oscillator was then varied over the course of the timeseries to
give the required shifts. We averaged many thousands of independent
realizations, for computations made at each of the shifts indicated
above, to recover estimates of the underlying profiles that agreed
excellently with the analytical predictions.


At first glance it would seem that the profile given by
equation~\ref{eq:linmeanprofile} is quite different from a
Lorentzian. However, for $\epsilon \ll 1$ the function simplifies to
give a Lorentzian. In this case, by solving analytically the
likelihood maximization of the profile described by
equation~\ref{eq:linmeanprofile}, it is possible to show that $H^*$
and $\Gamma^*$, the height and linewidth of this simplified
(Lorentzian) function are, to the first non-null order, related to the
original $H$ and $\Gamma$ by:
\begin{equation}
\label{eq:height} H^* = H ~ (1 - \epsilon^2 \frac{\pi}{6W}),
\end{equation}
and
\begin{equation}
\label{eq:linewdith} \Gamma^* = {\Gamma} ~ (1+\epsilon^2
\frac{\pi}{6W}).
\end{equation}
Here $W$ is the width in frequency (in units of the linewidth
$\Gamma$) over which the fit is made in the power frequency
spectrum. From these equations we can say that the linewidth will be
slightly overestimated, and the height slightly underestimated, if a
simple Lorentzian fitting model is used to fit the mode peaks. It is
interesting to note that the estimates $H^*$ and $\Gamma^*$ tend
towards $H$ and $\Gamma$ as $W$ increases. This means that ideally
with a sufficiently large frequency window it should be possible to
recover the original parameters of the mode to good accuracy.

Finally in this section, we note that the observed solar low-$l$
p-mode peaks are slightly asymmetric in shape, albeit at the level of
only a few per cent at most. It is also possible to derive a version
of equation~\ref{eq:linmeanprofile} based on an unperturbed profile
that is asymmetric, e.g., based on the widely-used asymmetric
formalism of Nigam \& Kosovichev (1998). The profile that is given is:
\begin{eqnarray}
\label{eq:linmeanprofileasym}
\left<P(\nu) \right> = & \frac{H}{2\epsilon} {\rm atan}
\left( \frac{2 \epsilon}{1 - \epsilon^2 + X^2}\right) \\
&+ B\frac{H}{2\epsilon} {\rm \log} \left( \frac{1 +\left( X +
\epsilon \right)^2} {1 +\left( X - \epsilon \right)^2} \right),
\end{eqnarray}
where $B$ is the peak asymmetry parameter. In the limit of small
frequency shifts we then have
\begin{equation}
\label{eq:approxmeanprofile} \left<P(\nu)\right> = \frac{H}{1 -
\epsilon^2  +X^2}\left[1 + 2BX\left(1+\epsilon^2
\eta(X,\epsilon)\right)\right], \label{eq:simpnk}
\end{equation}
with $\eta$ a function of $X$ and $\epsilon$ that satisfies
$\left|\eta(X,\epsilon)\right|\le 1$. Equation~\ref{eq:simpnk} shows
that the usual Nigam-Kosovichev profile still holds. However, as for
the height and the linewidth parameters the asymmetry will be changed
by a small amount, this amount being proportional to $\epsilon^2$.

 \subsection{Cosinusoidal variation in time}
 \label{sec:cos}

The linear model above is useful for representing observations made on
the steepest parts of the rising or falling phases of the solar
activity cycle, where the global activity varies approximately
linearly in time. But what if the observations also cover the other,
non-linear parts? We therefore also consider the following time
dependence for the mode frequency
 \begin{equation}
 \label{eq:solmodefreq}
 \nu(t) = {\nu_0} + \frac{\Delta\nu}{2}
 \left[ 1 - \cos\left(\frac{2\pi t}{P_{\rm
 cyc}}\right) \right],
 \end{equation}
because it mimics the `periodic' pattern of the solar activity
cycle. The variation is formulated as shown above so that $\nu_0$ is
again the unperturbed frequency (i.e., the frequency at minimum
activity); while $\Delta \nu$ is now the full amplitude (from minimum
to maximum) of the cyclic frequency shift (not the \emph{total} shift,
as in the linear model).  When the length of observation, $T$, equals
the cycle period $P_{\rm cyc}$ (or one-half of the period) it can be
shown that the average profile resulting from
equation~\ref{eq:solmodefreq} is:
 \begin{equation}
 \label{eq:cosmeanprofile}
 \left<P(\nu) \right> = \frac{\sqrt{L_1.L_2}} {\sqrt{1-\epsilon^2.F(L_1,L_2)}},
 \end{equation}
where
\begin{eqnarray*}
\label{eq:cosmeanprofileextra}
F(L_1,L_2) & = & \frac{4 L_1 L_2}{H\left(\sqrt{L_1}+\sqrt{L_2}\right)^2},\\
L_1 & = & \frac{H}{1+\left(X-\epsilon\right)^2},\\
L_2 & = &\frac{H}{1+\left(X+\epsilon\right)^2},\\
X &= & \frac{\nu - (\nu_0+\frac{\Delta\nu}{2})} {\Gamma/2},\\
\epsilon & = & \frac{\Delta \nu}{\Gamma}.
\end{eqnarray*}
In this case it is interesting to note that since the frequency spends
more time around its maximum and minimum values, power near these
extreme frequencies will have more weight in the time-averaged
profile, giving the profile a double-humped appearance. This is
reflected in the profile analytical expression through the two
Lorentzians $L_1$ and $L_2$.  It is obvious from
equation~\ref{eq:cosmeanprofile} that when $\epsilon \ll 1$ the
profile tends to a single Lorentzian.

\begin{figure}
 \centerline{\epsfxsize=8cm\epsfbox{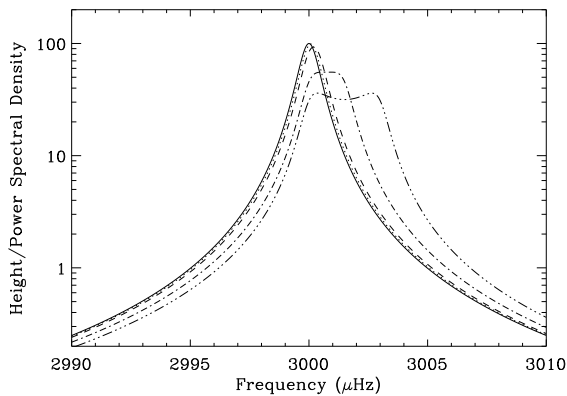}}

 \caption{Peak profiles expected for a single mode of width $1\,\rm
 \mu Hz$ in the power frequency spectrum of a time series within which
 the frequency was varied in a cosinusoidal manner. The timeseries is
 assumed to have length equal to one cycle period; while the amplitude
 of the cycle is $\Delta\nu/2$ (giving a total minimum-to-maximum
 shift in frequency of $\Delta\nu$).  Various linestyles are: no shift
 (solid line); $\Delta\nu=0.15\,\rm \mu Hz$ (dotted line); $0.40\,\rm
 \mu Hz$ (dashed line); $1.50\,\rm \mu Hz$ (dot-dashed line); and
 $3.0\,\rm \mu Hz$ (dot-dot-dot-dashed line). Note that only at the
 two largest shifts do the profiles depart significantly from the
 unperturbed (Lorentzian) profile.}

 \label{fig:cosprofs}
 \end{figure}

Figure~\ref{fig:cosprofs} shows predicted profiles from
equation~\ref{eq:cosmeanprofile}, assuming observations made over a
complete activity cycle, and with the same shifts $\Delta\nu$ that
were applied in the linear-model case (see figure~\ref{fig:profs} and
Section~\ref{sec:lin}). These profiles have again been validated by
averaging many independent power frequency spectrum realizations made
from stochastic harmonic oscillator timeseries.  As for the simpler
linear variation, it is only at the two largest $\Delta\nu$ that the
profiles depart appreciably from the unperturbed (Lorentzian) form,
here showing the predicted `humps' at the extreme frequencies of the
cycle. However, closer inspection again reveals some minimal
distortion of the Lorentzian shapes at the small Sun-like shifts. For
a given shift, this distortion appears to be slightly larger than in
the simpler, linear case. The implications of this distortion for
parameter bias are discussed in Section~\ref{sec:bias} below.


As in the case of a linear frequency shift it is also possible to
derive a version of equation~\ref{eq:cosmeanprofile} based on an
unperturbed profile that is asymmetric in shape. The profile that is
given is:
 \begin{equation}
 \label{eq:cosmeanprofileasym}
 \left<P(\nu) \right> = \frac{\sqrt{L_1.L_2}}
	{\sqrt{1-\epsilon^2.F(L_1,L_2)}}
 \left[1+2BX\left(1-\epsilon^2F(L_1,L_2)\right)\right],
 \end{equation}
where $B$ is again the peak asymmetry parameter.

\section{Estimation of frequency shift using new formalism}
\label{sec:est}

Solar-cycle frequency shifts of p modes are usually estimated by
extracting estimates of mode frequencies from short timeseries,
distributed at different epochs along the solar activity cycle. In
this section, our aim is to see whether it is instead possible to use
the new analytical expressions to extract estimates of the frequency
shifts. The observed time variations of the mode frequencies have a
qualitatively similar form to the cosinusoidal time variation on which
equations~\ref{eq:cosmeanprofile} and~\ref{eq:cosmeanprofileasym} are
based (e.g., see Chaplin et al. 2007a). The idea is that by using the
new equations to model the mode peaks in the peak-bagging codes it may
be possible to estimate the frequency shifts, since those shifts may
give rise to measurable distortions.

In our tests we assume the observations span approximately one
complete solar activity cycle, and so we therefore use fitting models
based on the cosinusoidal model. A cosinusoidal variation is of course
not an ideal match to the real observed time variation of the solar
activity cycle; however, our point here is to test the principle of
the technique. We begin with tests on artificial data that do have an
underlying variation of the frequencies that is cosinusoidal in
time. We then apply the technique to real low-$l$ timeseries, which
have lengths spanning approximately one 11-yr solar activity cycle.

\subsection{Application to artificial data}
\label{sec:appart}

We first made 1000 realizations of an artificial timeseries comprising
an $l=0$/2 mode pair, in which the modes were given underlying
parameters expected for low-$l$ modes at $\sim 3000\,\rm \mu
Hz$. Frequencies of the artificial modes were varied over time in a
cosinusoidal manner. The timeseries were each $T=11\,\rm yr$ long --
corresponding to the length of the artificial cycle, $P_{\rm cyc}$ --
and all modes were given a total frequency shift, from the minimum to
the maximum of the cycle, of $\Delta \nu = 0.4\,\rm \mu Hz$. The
artificial mode pairs were then fitted in power frequency spectra of
the timeseries to fitting models based on
equation~\ref{eq:cosmeanprofileasym}. The results showed it was
possible to extract estimates of the frequency shift given to the
modes, to a typical precision of $\sim 0.15\,\rm \mu Hz$.

In 50 cases out of the total of 1000 simulations (i.e., in 5\,per cent
of the realizations) the fits `locked onto' a null, or zero-valued,
estimate of the frequency shift. This is a recurrent problem when fits
are made for parameters which are too sensitive to the realization
noise (e.g., estimation of component frequency splittings of blended
modes).  In such cases the maximum of the likelihood function can be
far enough from the solution that it lies outside the range accessible
to the parameter.  The fit is therefore `stopped' by the hard limit,
which is zero in the case of the frequency shift.

As noted in Section~\ref{sec:est} above, the classic approach to
estimation of the shifts involves dividing the full timeseries into
shorter subseries, which are then fitted to yield time-dependent
estimates of the frequencies. Analysis of the resulting set of
frequency estimates yields an estimate of the total frequency
shift. When we applied this classic approach to the 1000 artificial
timeseries, we found it was possible to estimate the shift to a
precision of $\sim 0.07\,\rm \mu Hz$. The precision is clearly
superior to that given by fitting the new equations to the peak
profiles. This is not surprising: We showed in Section~\ref{sec:cos}
that for realistic low-$l$ frequency shifts the distortion of the mode
profiles is very modest. This makes it hard to measure the distortion,
and extract a robust estimate of the frequency shift, using the new
technique.

\begin{figure}
 \centerline{\epsfxsize=8cm\epsfbox{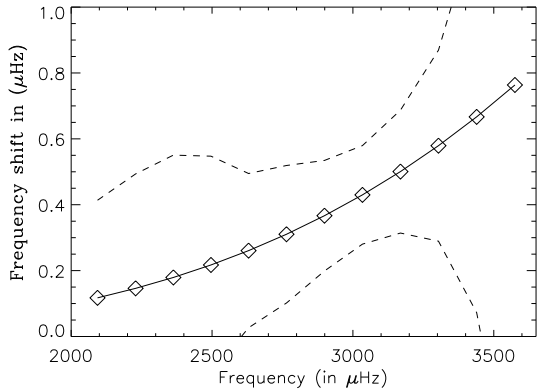}}

 \caption{Fitted frequency shifts (symbols, with associated $1\sigma$
 error envelope shown by dashed lines) and input frequency shifts
 (solid line) for modes in artificial $l=0$/2 mode pairs. The
 frequency shifts were estimated by fitting the artificial mode peaks
 to fitting models based on equation~\ref{eq:cosmeanprofileasym}.}

 \label{fig:freqshiftaccuracy}
 \end{figure}

We then extended the Monte-Carlo tests to simulate a range of mode
pairs across the low-$l$ power frequency spectrum. We adopted the
strategy of Toutain, Elsworth \& Chaplin (2005), whereby the
artificial underlying limit power frequency spectrum was computed and
then fitted. This strategy saves on computing time, since one does not
have to generate, and then fit, a large number of independent
timeseries to give useable
statistics. Equation~\ref{eq:cosmeanprofileasym} was used to make the
mode profiles.  The best-fitting uncertainties on fits made to these
artificial data gave direct estimates of the precision expected from a
single timeseries realization of the same length as that used to
compute the underlying limit power frequency spectrum. (Note the fits
recover the input bias accurately, since the fitting model was based
on the equations that were used to describe the artificial profiles.)
The results, which are plotted in figure~\ref{fig:freqshiftaccuracy},
show that the precision in the estimates is quite modest, particularly
at lower frequencies where the input frequency shifts are smallest in
size.


\subsection{Application to real data}
\label{sec:appreal}

Next, we made use of four timeseries of real disc-integrated Doppler
velocity observations of the Sun. Two timeseries comprised
Sun-as-a-star observations: one made of data collected by the
ground-based BiSON between 1993 and 2003; and one made of data
collected by the GOLF instrument on board the \textit{ESA/NASA} SOHO
spacecraft between 1996 and 2004. The other two timeseries comprised
resolved-Sun observations that were spatially averaged to give a
Sun-as-a-star proxy signal: one made of data collected by the
ground-based GONG between 1995 and 2004; and one made of data
collected by the MDI instrument on SOHO between 1996 and 2006.  Each
of the four timeseries covers more or less one 11-yr cycle of solar
activity.

The power frequency spectrum of each timeseries was computed, and
low-$l$ mode pairs were fitted to models based on
equation~\ref{eq:cosmeanprofileasym}. We used
equation~\ref{eq:cosmeanprofileasym}, rather than
equation~\ref{eq:cosmeanprofile}, because the real low-$l$ peaks show
small amounts of asymmetry. The $l=0$/2 pairs were fitted assuming all
constituent components had the same frequency shift. For the $l=1$/3
pairs, our fitting results demonstrated that the relative weakness of
the $l=3$ peaks meant the distortion of their profiles could not be
fitted reliably. At $l=1$, the blending of the constituent components
of each mode caused some problems for the fitting, which was
manifested by cross-talk between the fitted frequency shift and frequency
splitting parameters.

\begin{figure}
 \centerline{\epsfxsize=8cm\epsfbox{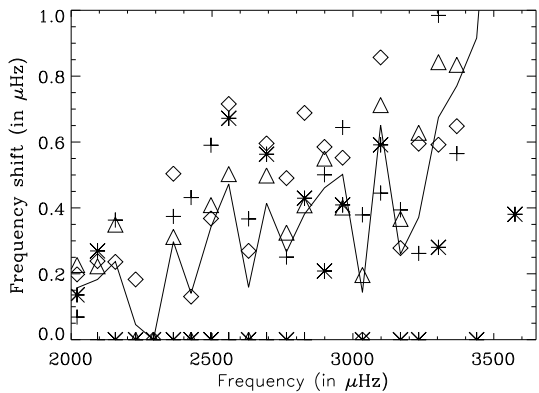}}

 \caption{Estimated frequency shifts, averaged over $l=0$, 1 and 2
 modes, as extracted from fits to BiSON (stars), MDI (crosses) GONG
 (diamonds) and GOLF (triangles) power frequency spectra. The solid
 line is an average of all four sets of frequency shifts. The
 frequency shifts were estimated by fitting the artificial mode peaks
 to fitting models based on equation~\ref{eq:cosmeanprofileasym}.}

 \label{fig:freqshift}
 \end{figure}

Figure~\ref{fig:freqshift} shows the estimated frequency shifts as a
function of mode frequency, averaged over $l=0$, 1 and 2 for each
timeseries (see caption).  As we would have expected, given the
simulation results discussed in the previous section, it was not
always possible to extract accurate estimates of the frequency shifts,
and there were therefore several occasions on which the fits locked
onto zero-valued estimates.  Nevertheless the non-zero best-fitting
values follow the expected trend in frequency, giving estimates of the
shifts that are consistent with previous frequency-dependent estimates
for the low-$l$ modes (e.g., Chaplin et al. 2001; Jim\'enez-Reyes et
al. 2001; Gelly et al. 2002), apart from the highest frequencies,
where the shifts are somewhat larger.


\section{Bias on mode parameters}
\label{sec:bias}

In the previous sections we have seen that mode peaks of the low-$l$
solar p modes should be slightly distorted, with respect to the basic
Lorentzian-like profiles, when they are observed in power frequency
spectra made from observations spanning a substantial fraction or more
of the 11-yr solar activity cycle. This distortion is caused by
variation of the mode frequencies over time. Even though the
distortion is modest we have demonstrated that it is possible to
measure the distortion (and estimate frequency shifts) by using new
fitting models.

This result raises an important question: if the mode profiles are
indeed distorted, what bias might we expect in the best-fitting mode
parameters were we to continue to use the (inaccurate) Lorentzian-like
fitting models (e.g., the asymmetric Nigam-Kosovichev formalism)? We
use results on artificial (Section~\ref{sec:appart}) and real
(Section~\ref{sec:appreal}) solar p-mode data to seek an answer to
this question.

\begin{figure*}
 \centerline{\epsfxsize=8cm\epsfbox{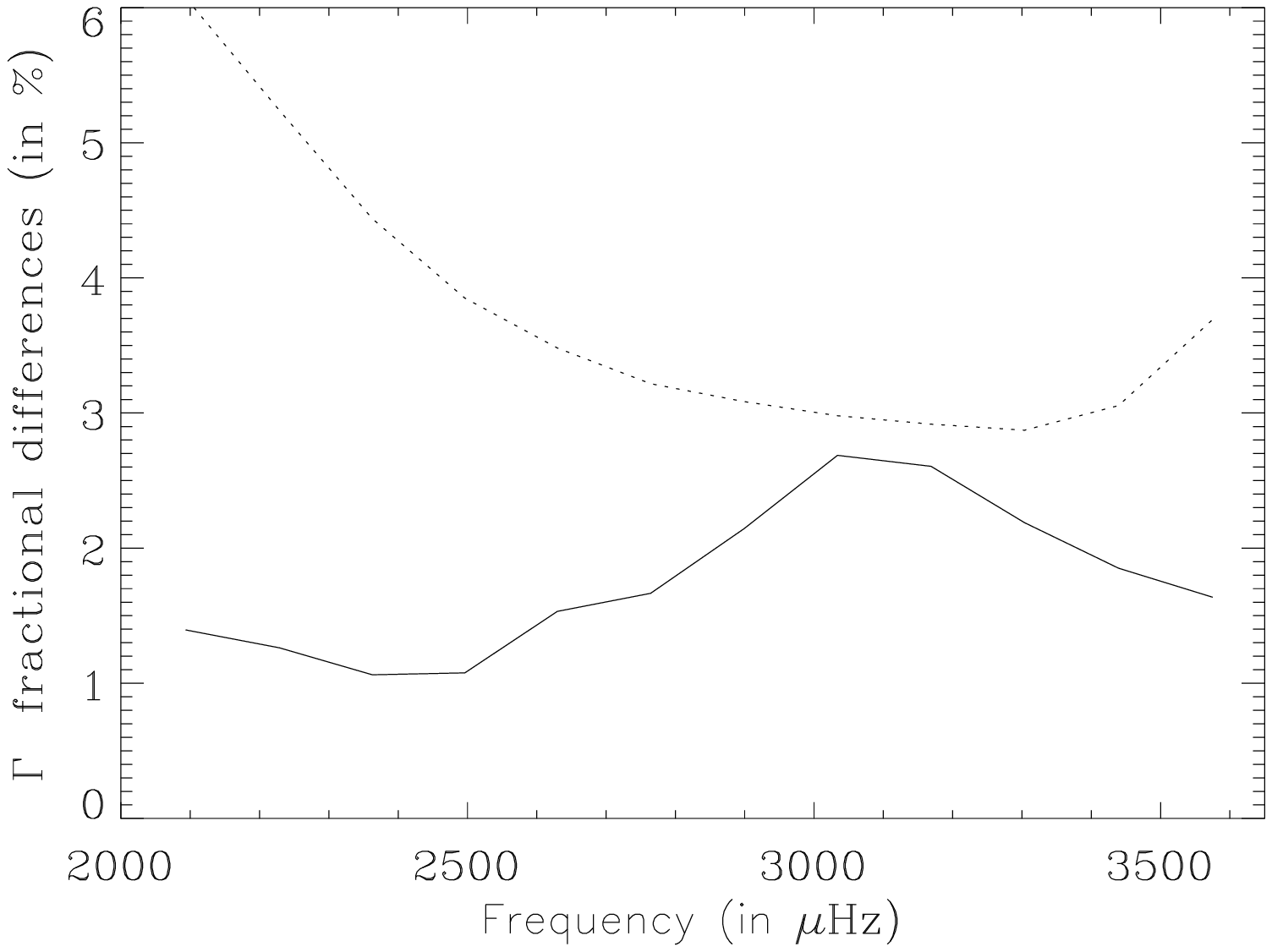}
             \epsfxsize=8cm\epsfbox{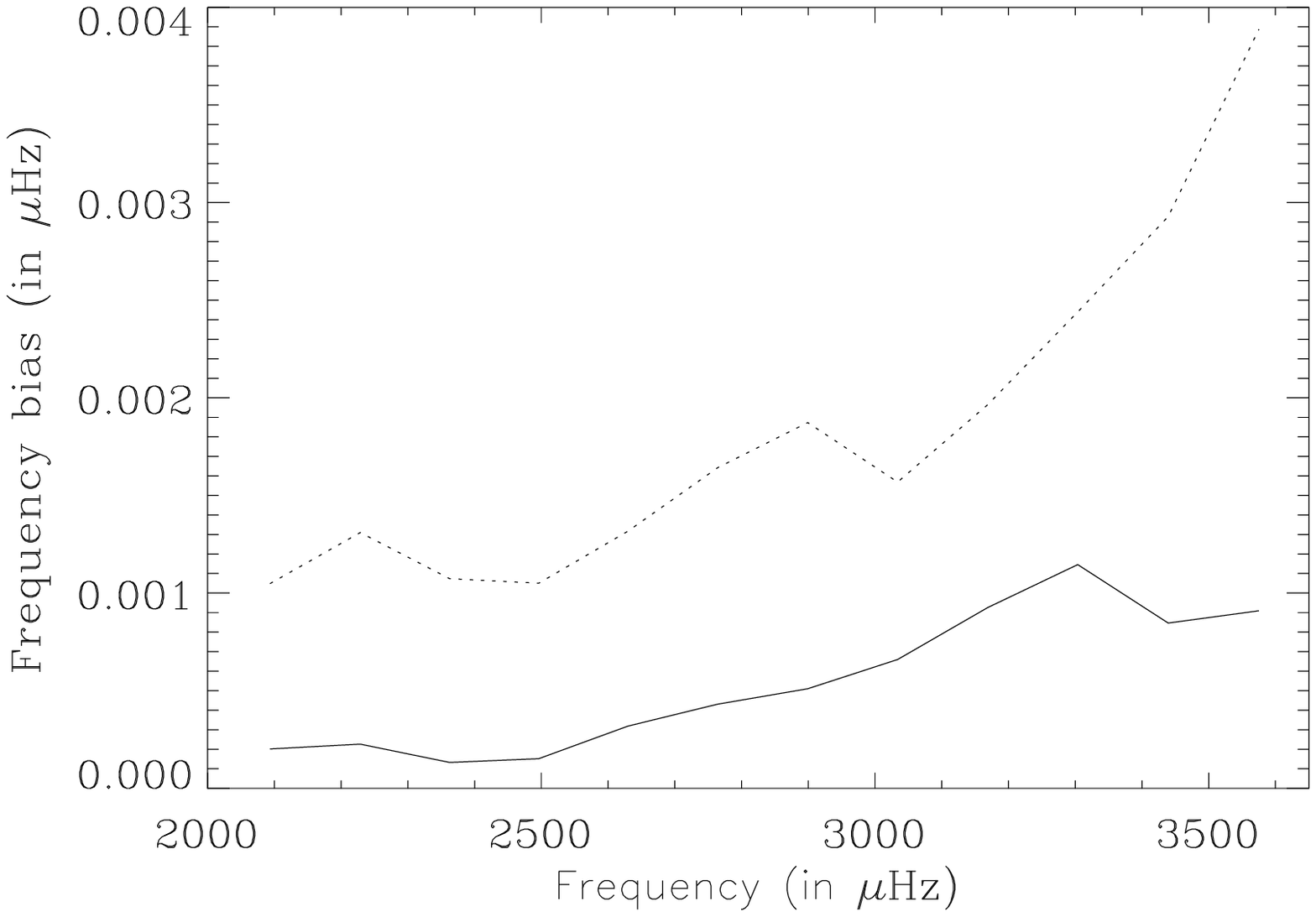}}

 \caption{Expected linewidth (left-hand panel) and frequency
 (right-hand panel) parameter bias given by fitting the usual
 asymmetric Lorentzian-like fitting models to the mode peaks, models
 which do not account for the distortion of the underlying
 profiles. The dotted line in the left-hand panel shows the full
 best-fitting linewidth uncertainty. The dotted line in the right-hand
 panel is one tenth the size of the formal frequency uncertainty.}

 \label{fig:freqMonteCarlo}
 \end{figure*}

Let us consider first results on artificial data. We again adopted the
strategy of Toutain, Elsworth \& Chaplin (2005), creating artificial
underlying limit power frequency spectra corresponding to 11-yr of
observations.  Equation~\ref{eq:cosmeanprofileasym} was used to make
the mode profiles, with the assumed input frequency shifts having the
same sizes as those shown in figure~\ref{fig:freqshiftaccuracy}. The
low-$l$ pairs were then fitted to fitting models made with the
Nigam-Kosovichev formalism. Comparison of the fitted and input
parameters gave the bias estimates plotted as solid lines in the
left-hand (for linewidth) and right-hand (for frequency) panels of
figure~\ref{fig:freqMonteCarlo}. 


The dotted line in the left-hand panel shows the full best-fitting
linewidth uncertainty (which is of similar magnitude to the height
uncertainty). The overestimation of the linewidth -- a similar-sized
underestimation of the height is also observed -- is seen to be of a
size comparable to the estimated uncertainties. The dotted line in the
right-hand panel is one-tenth the size of the best-fitting mode
frequency uncertainty.  Bias in the mode frequencies is evidently very
small indeed, in contrast to the linewidth parameter. That said, we
need to interpret the frequency result with a little care.

When the observations span either a complete cycle (as here), or one
half of a cycle, the frequency parameter should in principle not be
biased because the distortion is symmetric in frequency. Evidently,
the small residual bias exhibited in the right-hand panel of
figure~\ref{fig:freqMonteCarlo} is due to other effects (e.g., from
parameter cross-talk in the fitting, and the impact of the asymmetric
shapes of the peaks). We would, however, expect there to be a bias
from the distortion effect when observations span, say, one-quarter of
a cycle. Under these circumstances the distortion will not be
symmetric in frequency.

\begin{figure}
 \centerline{\epsfxsize=8cm\epsfbox{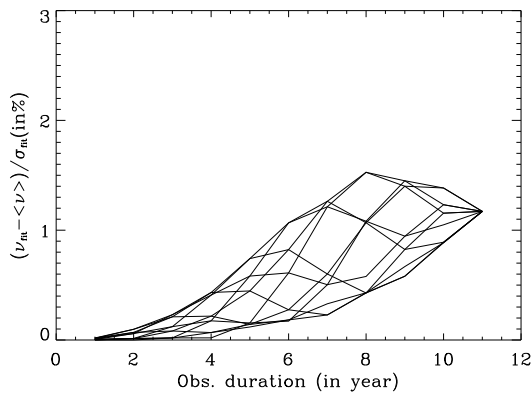}}

 \caption{Estimated frequency bias (from simulations) for modes at
 $3000\,\rm \mu Hz$. The bias is plotted in units of the formal
 frequency uncertainties as a function of the length of the
 observations. Each of the curves show results for different starting
 points along the simulated 11-yr solar activity cycle.}

 \label{fig:lotsof}
 \end{figure}

To test this case, and other intermediate cases, we made further
artificial datasets. Artificial power frequency spectra were made to
mimic observations ranging in length from 1 to 11\,yr. Furthermore,
the observations were assumed to start anywhere from the beginning of
the activity cycle up to year 10 of the cycle (in 1\,yr
steps). Estimates of the frequency bias of modes at $3000\,\rm \mu
Hz$, as a function of the simulated length of the observations, are
plotted in figure~\ref{fig:lotsof}. The various curves show the bias
-- as a percentage of the formal frequency uncertainty -- for
different starting points along the 11-yr cycle. The magnitude of the
bias never rises above a few per cent of the frequency uncertainty. We
may therefore conclude that frequency bias is not a major cause for
concern.


\begin{figure*}
 \centerline{\epsfxsize=8cm\epsfbox{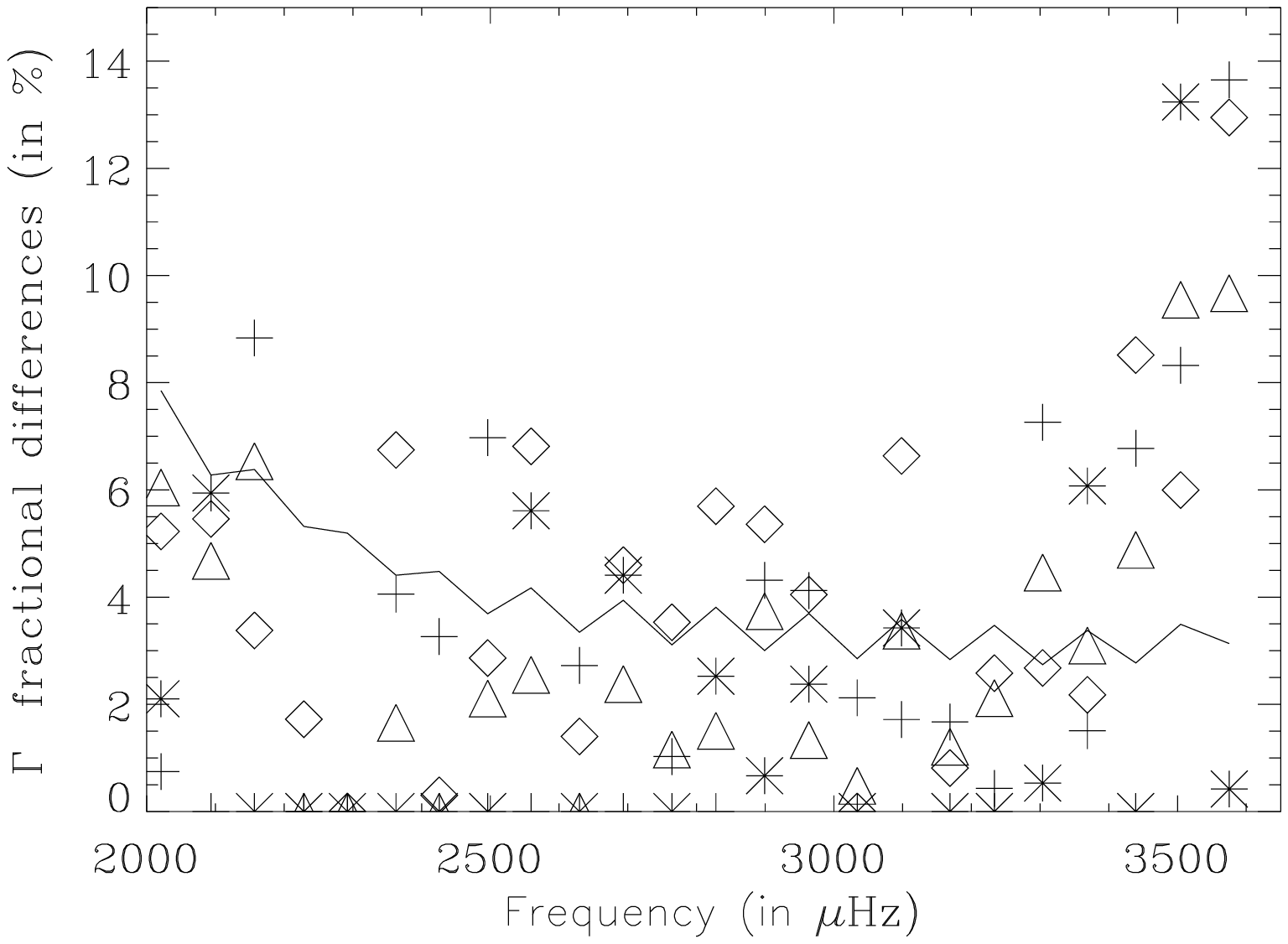}
             \epsfxsize=8cm\epsfbox{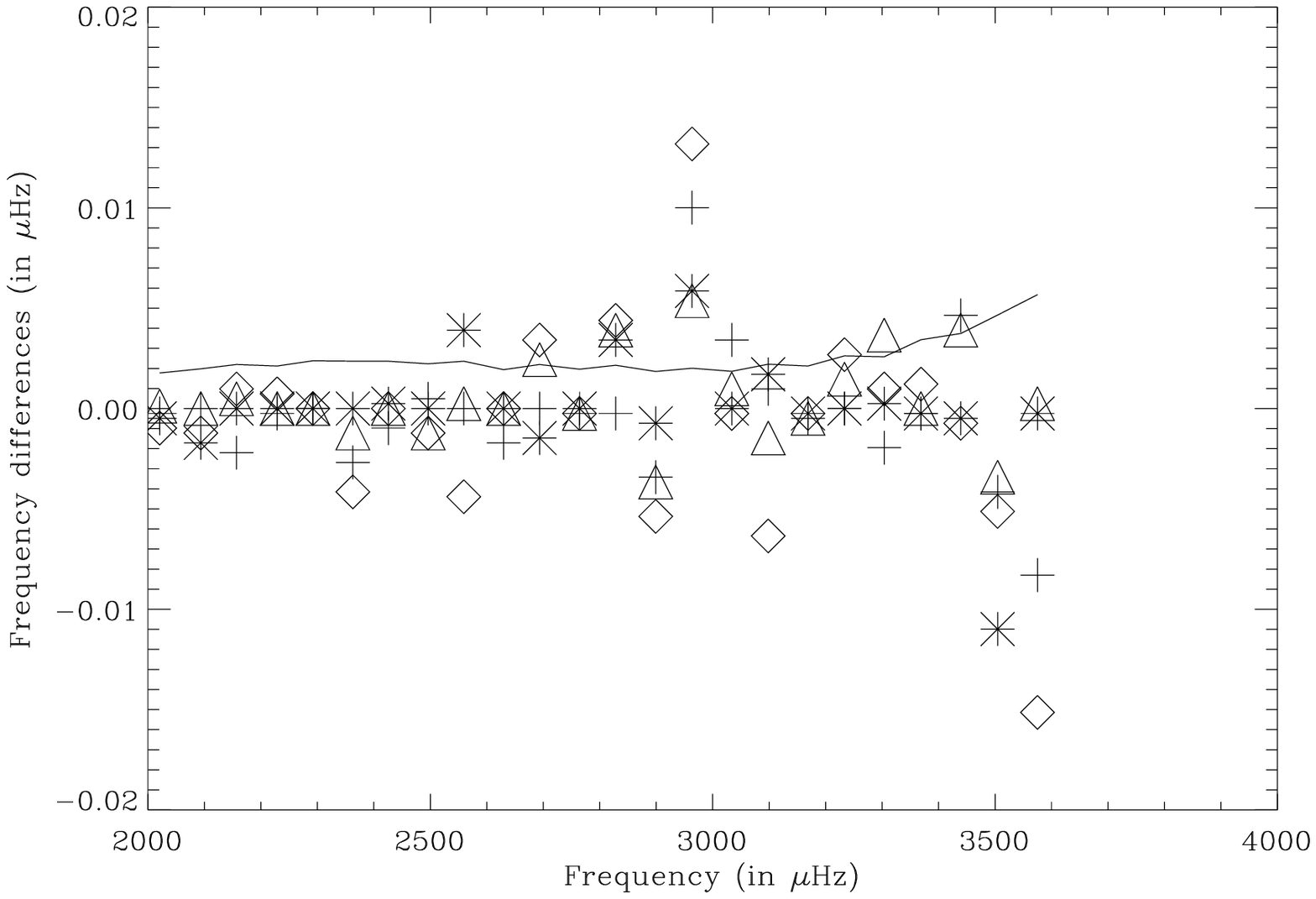}}

 \caption{Differences between results of fitting power frequency
 spectra with fitting models based on the Lorentzian-like Nigam \&
 Kosovichev model and equation~\ref{eq:cosmeanprofileasym}. The
 left-hand panel shows linewidth differences, the right-hand panel
 frequency differences (averaged over $l=0$, 1 and 2 modes) in BiSON
 (stars), MDI (crosses) GONG (diamonds) and GOLF (triangles) frequency
 power spectra. The solid line in the left-hand panel shows the
 typical best-fitting linewidth uncertainty; while the solid lines in
 the right-hand panel correspond to plus and minus one-tenth of the
 typical formal frequency uncertainty.}

 \label{fig:freqdiffs}
 \end{figure*}

Next, we fitted low-$l$ modes in the BiSON, GOLF, GONG and MDI power
frequency spectra (Section~\ref{sec:appreal}) to fitting models made
with the Nigam-Kosovichev formalism. By taking differences between
these results and those from fits to models based on
equation~\ref{eq:cosmeanprofileasym} we had another means of judging
the bias. The results (symbols for different timeseries) are shown in
the two panels of figure~\ref{fig:freqdiffs}. The solid line in the
left-hand panel shows the full best-fitting linewidth uncertainty;
while the solid lines in the right-hand panel correspond to plus and
minus one-tenth of the formal frequency uncertainty.


The results are shown to be in reasonable agreement with those of the
artificial data (figure~\ref{fig:freqMonteCarlo}). Overestimation of
the linewidths (left-hand panel of figure~\ref{fig:freqdiffs}) is on
average larger than the overestimation implied by the results of the
simulations. We believe this largely reflects the fact that in the
real data it is not only the frequencies that vary over the solar
cycle, but also the heights, linewidths (e.g., Chaplin et al. 2000)
and peak asymmetries (Jim\'enez-Reyes et al. 2007). In deriving the
equations (e.g., equation~\ref{eq:cosmeanprofileasym}), and
constructing the artificial data for the simulations, we assumed only
the frequencies varied in time. Proper allowance will be made for the
height, linewidth and asymmetry variations in the next phase of this
work.

Bias in the frequencies (right-hand panel of
figure~\ref{fig:freqdiffs}) is again shown to be but a small fraction
of the size of the frequency uncertainties.

\section{Conclusion}
\label{sec:conc}

We have studied the impact of the solar-cycle frequency shifts on the
underlying shapes of p-mode peaks seen in power frequency spectra made
from data spanning large fractions of the cycle period. Analytical
descriptions of the resulting mode profiles were presented for two
functions describing the shifts in time: a simple linear variation;
and a cosinusoidal variation to mimic the full solar activity cycle.

We presented plots of the profiles expected for shifts similar in size
to, and also larger than, those observed for the low-$l$ solar p
modes. The analytical predictions were also validated by Monte Carlo
simulations with artificial data. In summary, we showed that while any
distortion of the Lorentzian-like profiles of the solar p modes is
very modest, it is nevertheless just detectable. Furthermore our
analysis indicates that by fitting modes to the usual Lorentzian-like
models -- which do not allow for the distortion -- rather than new
models we derive, overestimation (underestimation) of the linewidth
(height) parameter results. This bias is estimated to be of size
comparable to the observational uncertainties given by datasets of
length several years. Bias in the frequency parameter is much less of
an issue, being over an order of magnitude smaller than the frequency
uncertainties.

The distortion discussed in this paper may of course be an issue for
analysis of multi-year asteroseismic datasets on some stars that show
Sun-like oscillations. The effect will be most important in those
stars for which the ratio of the stellar-cycle frequency shifts to the
mode linewidths is larger than for the Sun. Indeed, visible distortion
of the mode profiles in asteroseismic data may provide an initial
diagnostic of strong stellar-cycle signatures over the duration of the
observations (Chaplin et al. 2007b; Metcalfe et al. 2007).

We finish by offering an answer to the question posed by the title of
this paper. As far as the low-$l$ solar p-mode frequencies are
concerned, there is probably no need to worry about the distortion
introduced by the frequency-shift effect. However, in the case of the
height and linewidth parameters, a systematic bias of $1\sigma$ or
more means we do need to worry if we wish to obtain accurate estimates
of these parameters.

 \section*{ACKNOWLEDGMENTS}

BiSON is funded by the UK Science and Technology Facilities Council
(STFC).  We would like to thank all those who are, or have been,
involved in BiSON. GOLF and MDI and are the result of the cooperative
endeavours of several institutes, to whom we are deeply indebted. SOHO
is a mission of international cooperation between \textit{ESA} and
\textit{NASA}.  This work also utilizes data obtained by the Global
Oscillation Network Group (GONG) Program, managed by the National
Solar Observatory, which is operated by AURA, Inc. under a cooperative
agreement with the National Science Foundation. The data were acquired
by instruments operated by the Big Bear Solar Observatory, High
Altitude Observatory, Learmonth Solar Observatory, Udaipur Solar
Observatory, Instituto de Astrof\'isica de Canarias, and Cerro Tololo
Interamerican Observatory. The authors acknowledge the support of
STFC.

\end{document}